\documentclass[smallextended]{svjour3}       
\smartqed  
\usepackage{graphicx}
\usepackage{amssymb}
\usepackage{siunitx}
\usepackage{mathtools}
\usepackage{algorithmic}
\usepackage{algorithm}
\usepackage{hyperref}
\usepackage{url}
\usepackage{float}
\usepackage{color}
\usepackage[dvipsnames]{xcolor} 
\usepackage{caption}
\usepackage{subcaption}
\usepackage{placeins}
\begin{document}

\title{Robust GPU-based Virtual Reality Simulation of Radio Frequency Ablations for Various Needle Geometries and Locations}
\thanks{DFG: MA 6791/1-1; Nvidia GPU grant 2018
}
\titlerunning{Robust GPU-based RFA-Simulation}        

\author{Niclas Kath$^*$        \and
        Heinz Handels       \and
        Andre Mastmeyer$^*$
}


\institute{A. Mastmeyer \at
              Institute of Medical Informatics, University of Luebeck, Germany \\
              Tel.: +49-451-3101-5608\\
              Fax: +49-451-3101-5604\\
              \email{mastmeyer@imi.uni-luebeck.de}             \\
              \emph{$^*$} Equal author contributions to this paper  
           \and
           N. Kath \at
              \email{niclas.kath@student.uni-luebeck.de}
}

\date{Received: date / Accepted: date}

\maketitle

\begin{abstract}

Purpose: Radio-frequency ablations play an important role in the therapy of malignant liver lesions. The navigation of a needle to the lesion poses a challenge for both the trainees and intervening physicians. 

Methods: This publication presents a new GPU-based, accurate method for the simulation of radio-frequency ablations for lesions at the needle tip in general and for an existing visuo-haptic 4D VR simulator. The method is implemented {real-time} capable with Nvidia CUDA.

Results: {It performs better than a literature method concerning the theoretical characteristic of monotonic convergence of the bioheat PDE and a \textit{in vitro} gold standard with significant improvements ($p<0.05$) in terms of Pearson correlations. It shows no failure modes or theoretically inconsistent individual simulation results after the initial phase of 10 seconds. On the Nvidia 1080 Ti GPU it achieves a very high frame rendering performance of $>$480 Hz.}

Conclusion: {Our method provides a more robust and safer real-time ablation planning and intraoperative guidance technique, especially avoiding the overestimation of the ablated tissue death zone, which is risky for the patient in terms of tumor recurrence. Future \textit{in vitro} measurements and optimization shall further improve the conservative estimate.}

\end{abstract}

\section{Introduction}
\label{intro}

In liver cancer therapy besides the conservatively used open laparotomy \cite{goldberg2002comparison,oshowo2003comparison}, there are several options for the treatment of primary liver tumors. 

\begin{figure}[htpb!]
\centering
\begin{subfigure}[b]{0.3\textwidth}
    \includegraphics[width=0.85\textwidth]{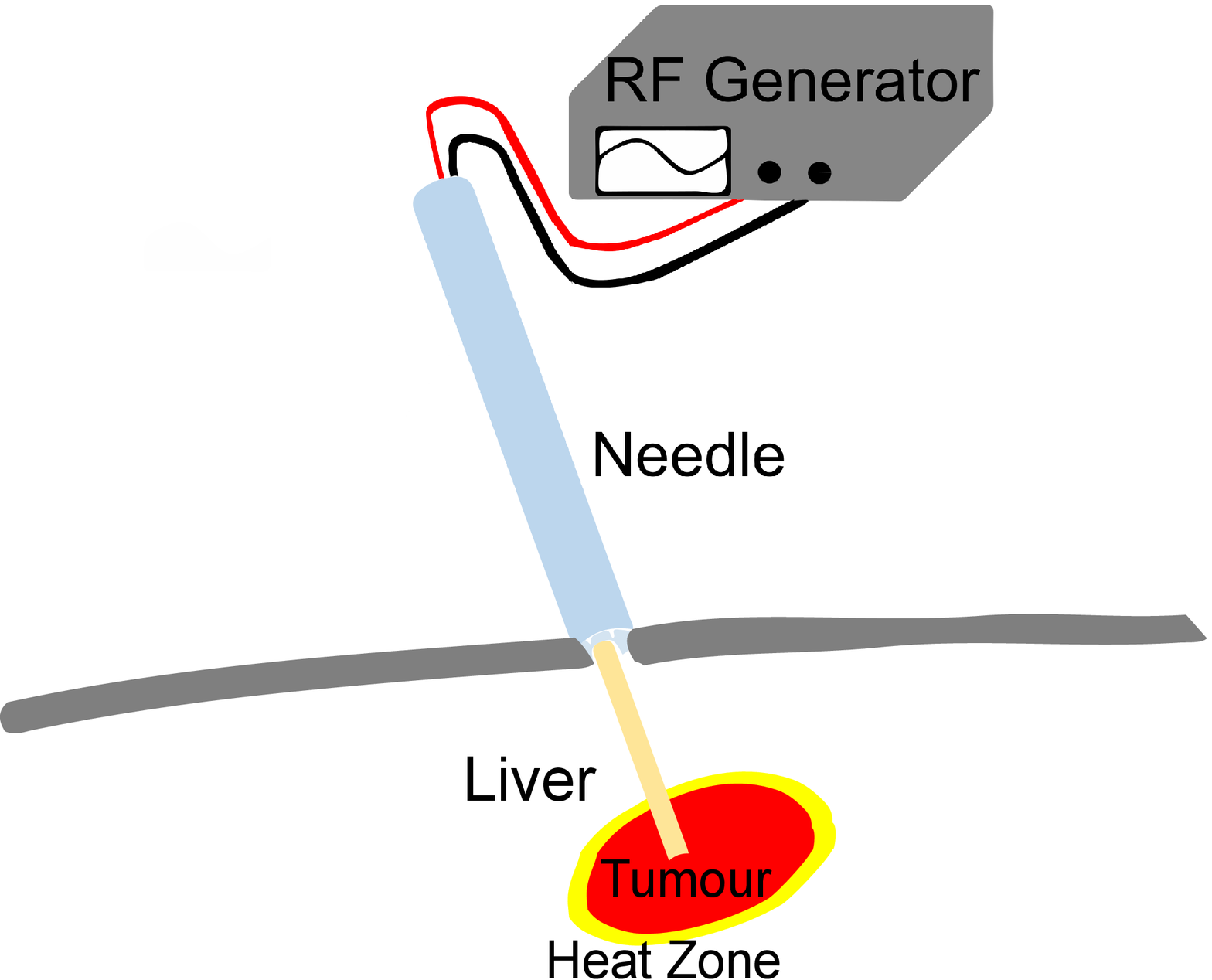}
    \caption{RFA setup}
    \label{RFAScheme}
\end{subfigure}
\begin{subfigure}[b]{0.3\textwidth}
    \includegraphics[width=.85\textwidth]{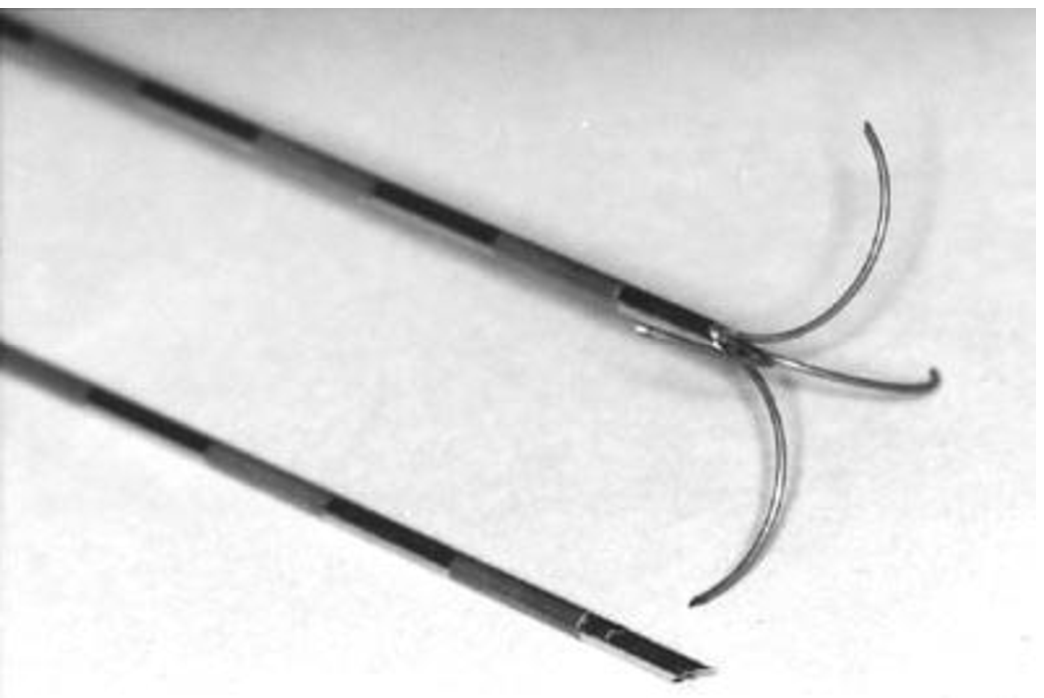}
    \caption{Needle with 4 tip extension wires (umbrella)}
    \label{fig:umbrella}
\end{subfigure}
\caption{(a) Scheme of a bipolar RFA. The heat zone (yellow, \SI{42,5}{\celsius}) has to outgrow the tumor (red) to prevent recurrence of the tumor. (b) Complex needle geometry: The wires are extended inside the tumor to cover bigger lesions with  ablation heat.}
\end{figure}

The oral endoscopic access is chosen for lesion in or at the main ducts (e.g. CHD). The minimally invasive  laparoscopic access (endoscopic keyhole surgery) is favorable for peripheral liver boundary tumors. 
For smaller tumors in the vicinity of hepatic arteries, the minimally invasive and lower-risk radio-frequency ablation (RFA) \cite{2675-01} is possible (Fig. \ref{RFAScheme}). RFA is the most common \cite{goldberg2002comparison} and geometrically flexible ablation technique (see Fig. \ref{fig:umbrella}) for focal tumors and therefore is the basis of this work. A needle containing an electrode is used to access the target, and the electrode is used to emit high frequency currents to the needle tip. Fluoroscopy or ultrasound \cite{2675-10} image guidance helps to navigate the tip to the target lesion. Compared to open laparotomy, the survival chance is only 2\% worse \cite{oshowo2003comparison}.
The long chained proteins of the cells disintegrate 
leading to the tissue ablation.
Clinically, total necrosis instantaneously occurs at temperatures below \SI{-40}{\celsius} (cryoablation) or exceeding \SI{60}{\celsius} for most cell types \cite{nikfarjam2005mechanisms}.

Uncommonly used, microwave ablation is a rather new technique which heats tissue around a coil in the tip to remove cancer tissue \cite{2675-24}. Radiation tip brachytherapy is more common in the lower abdomen targeting the cervix or prostate \cite{2675-15,mastmeyer2015model}.
The treatment of spherical or ellipsoidal tumours is faster, subject to shorter fluoroscopy exposition and more painless \cite{chen2015efficacy} with cryotherapy \cite{kuck2016cryoballoon}, which destroys tumour cells with ice crystals from quick freezing and thawing \cite{2675-25}.
Transarterial chemoembolization (TACE) is another treatment option (yet more side effects) for rather small or otherwise unresectable tumours extending the lifetime of patients \cite{llovet2003systematic}.
Laser ablation based on dispersed laser light from fibers is another safe and successful thermal ablation technique \cite{izzo2003other}.

The ablation principle is shown in Fig. \ref{RFAScheme}. Currents from a radio-freqency generator heat up the tissue around the needle tip.
Within the 4D VR needle ablation simulator from our group \cite{2675-08,2675-09,2675-10,2675-14,2675-16,2675-17}, we use an efficiently parallel calculable temperature model \cite{2675-02} specialized to liver lesions.
The new GPU-based algorithm also compares qualitatively (patient safety) favorable to the literature \cite{2675-05,2675-18} and theoretical characteristics. Keep in mind throughout the rest of the paper in RFA planning, overestimation of the heat zone can be deadly by finally leaving residual tumor cells (under-treatment)
and the 100\% reproducability of simulations (no statistical errors).
Currently, the training of ablations is carried out in a risky set-up directly on the living patient. However, VR training and planning simulations such as ours \cite{2675-13,2675-08} are on the rise and safe over-{real-time} ablation planning is highly relevant for intraoperative planning systems.

\section{Material and Methods}

The calculation of the heat distribution is based on virtual body models, which essentially consist of segmented 3D CT image data. To better found our argumentation, we first show some theoretical thoughts.

\subsection{Theoretical Background}

The fundamental solution of the second order hyperbolic partial differential heat equation (PDE) \cite{evans2012numerical}:

\begin{equation}
\dot T = k \nabla^2 T 
\end{equation}

describes the temperature $T$ distribution in a solid medium over time. $k$ is a constant describing the temperature (substance or energy) spread rate. Now, we can introduce a border condition with an initial point source $T(x,0)=g(x)=\delta(x)$ of heat at a known position, i.e. 0 with heat $\SI{1}{\celsius}$. This is known as the fundamental heat distribution problem solved by convolution.

The fundamental solution to the above problem is a typical Green's function:

\begin{equation} \label{eq-funda}
    T_f(x,t)=\frac{1}{\sqrt{4\pi kt}}\exp\left(-\frac{x^2}{4kt}\right).
\end{equation}
 
\subsubsection{General Formal Framework}
Generally, a linear combination of solutions of other well-known fundamental problem solutions subject to other elementary initial (IC, initial temperature distribution) or border conditions (BC, constant heat source) or slight modifications in the PDE term in Eq. \ref{eq-funda} on the right hand side (AT) \cite{evans2012numerical}:

\begin{equation}\label{eq:setup}
    \begin{cases} 
    \dot T=k\cdot \frac{\partial T}{\partial x^2}+f(x, t) & \textrm{additional terms $f$ (AT)}  
    \\ T(x,0)=g(x) & \textrm{initial condition (IC)} 
    \\ T(0,t)=h(t) & \textrm{boundary condition (BC)}
    \end{cases}
\end{equation}

where the first line is the fundamental Laplacian with other additive terms $f(x,t)$, IC corresponds to a initial temperature distribution (later: image) at time point zero and BC suits to represent a constant heat source $h(t)=T_0$. In this chapter for brevity, we set $g(x)=0$ and concentrate to solve the AT and BC problems for our needs.

For elaborating just the characteristics, we sum up the math: The solution with respect to the special terms, set here to $f(x,t)=0, g(x)=0, h(t)=T_0$, is a linear combination of appropriate Green's function solutions resulting from the linear operation of convolutions of the type shown in Eq. \ref{eq-funda} (Bell curves). 

This translates perfect to the characteristic simulation problem of needle ablation, e.g. using a needle tip heat source $T(0,t)=h(t)=\delta(x)\cdot\SI{90}{\celsius}$. 
The solution of the time convolution:

\begin{equation}
    T_{BC}(x,t)=\int_{0}^{t} \frac{x}{\sqrt{4\pi k(t-s)^3}} \exp\left(-\frac{x^2}{4k(t-s)}\right)h(s)\,ds, \qquad\forall x>0
\end{equation}

can be found using the indefinite integral - as $h(s)=T_0=const.$, i.e.:

\begin{equation}
   T_0\cdot \int \frac{x}{\sqrt{4\pi kt^3}} \exp\left(-\frac{x^2}{4kt}\right)\,dt 
   =  -T_0\cdot\frac{\sqrt{k}t^{3/2}\cdot \textrm{erf}\left(\frac{x}{2\sqrt{k}}\right)}{\sqrt{kt^3}}+\textrm{const.}
\end{equation}

and finally yields with integration limits applied and $const.=T_0$:

\begin{equation}
T_{BC}(x,t)=T_0-T_0\cdot \frac{\sqrt{k}t^{3/2}\cdot \textrm{erf}\left(\frac{x}{2\sqrt{k}}\right)}{\sqrt{kt^3}}
\end{equation}

\begin{figure}[htpb!]
\centering
\vspace{-1.2cm}
\begin{subfigure}[b]{0.4\textwidth}
    \includegraphics[width=\textwidth]{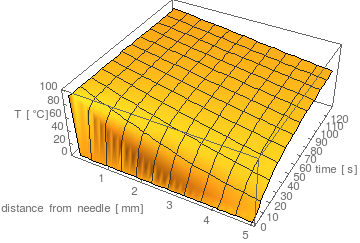}
    \caption{Space-time overview}
    \label{anaOverview}
\end{subfigure}\\
\begin{subfigure}[b]{0.3\textwidth}
    \includegraphics[width=\textwidth]{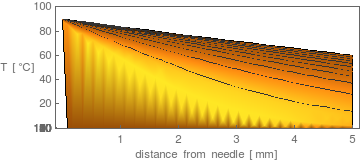}
    \caption{Space characteristic}
    \label{anaSpace}
\end{subfigure}
\begin{subfigure}[b]{0.3\textwidth}
    \includegraphics[width=\textwidth]{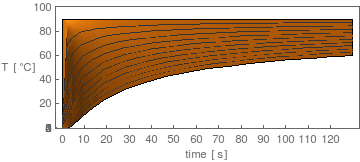}
    \caption{Time characteristic}
    \label{anaTime}
\end{subfigure}
\caption{The space-time characteristic of the analytical solution of the heat equation: Monotonic convergence in time and space: (a) 3D temperature space-time plot. (b) The final distribution of temperature over distance (radial from heat source) is a line. (c) Temperatures converge monotonically to that line over time.}
\end{figure}

This characteristic analytic solution $T=T_{BC}$ is plotted with $h(t)=T_0=\SI{90}{\celsius}$ and $k=0.52$ in Fig. \ref{anaOverview} over space $x\geq 0$ and time $t\geq 0$. The solution shows monotonic convergence of temperature (Figs. \ref{anaSpace}, \ref{anaTime}).

\subsubsection{Cell Decay Terms from the Comparison Work}\label{sec:tissState}
\begin{figure}[htpb!]
\centering
\vspace{-1.5cm}
    \includegraphics[width=.5\textwidth]{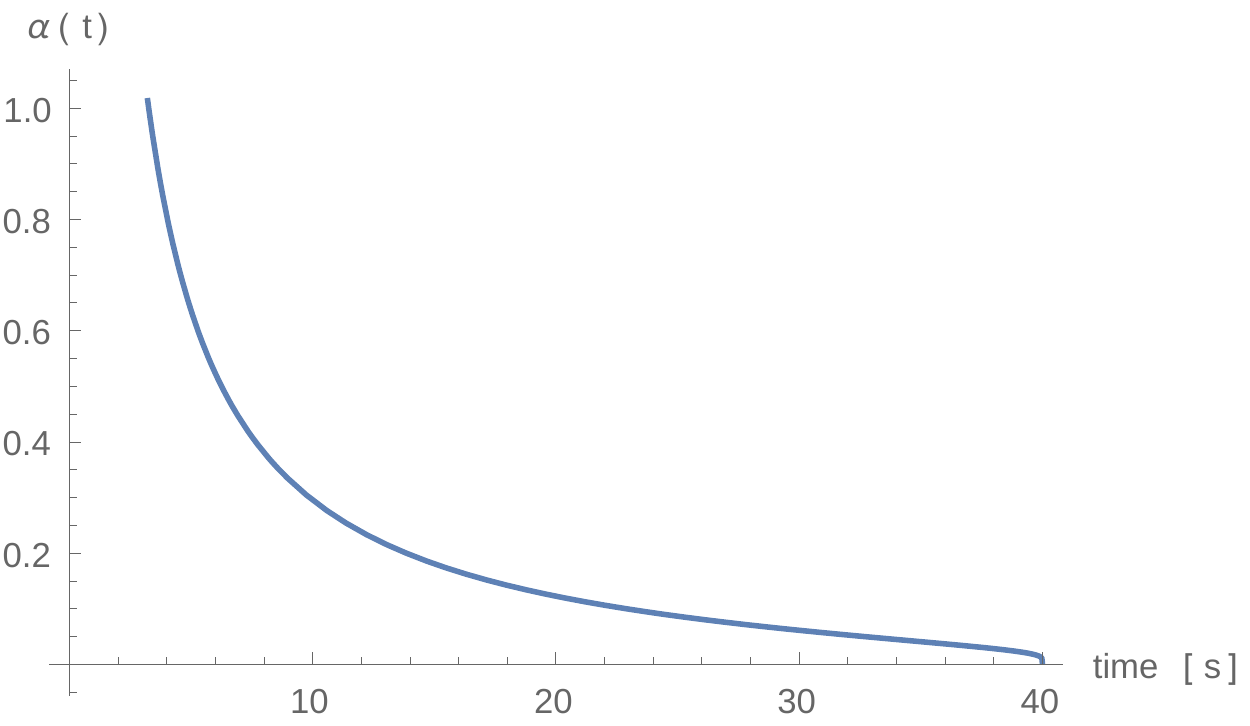}
    \caption{ Characteristical tissue decay $\alpha(t)$ with tissue death after \SI{40}{\second}.}
    \label{fig-Arrhenius}
\end{figure}

Linte et al. \cite{2675-05} proposed a tissue state decay term based on an inverted Arrhenius equation:

\begin{equation} \label{eq-tissState}
    \alpha(t)
    =\frac{1}{ln(c(0)/c(t))}
    =\left[A\int_0^t \textrm{exp}\left(-\frac{\Delta E}{RT(t)}\right)dt\right]^{-1}
\end{equation}

with symbols
$c(t)$ as the healthy cell concentration at time $t$,
$R$ as the universal gas constant (\SI[per-mode=fraction]{8.31}{\joule\per\mole\per\kelvin}), 
$A$ as a frequency factor for the kinetic expression (\si{\Hz}), 
$\Delta E$ denotes the activation energy for the irreversible cell damage reaction ($\si[per-mode=fraction]{\joule\per \mole}$) and 
$T$ denotes the tissue temperature (\si{\celsius}) \cite{2675-05,2675-20}:
Damaged tissue with no cooling by blood flow corresponds to $\alpha=0$ at some later time point, and fully functional tissue is represented by $\alpha=1$ at $t=0$ as shown in Fig. \ref{fig-Arrhenius}.
$\alpha$ is a monotonically decreasing term from 1 to 0 under-weighting the blood cooling.

\color{black}
\subsection{Specialized Methods for Liver Ablation Simulation}
The segmented regions of the virtual bodies \cite{2675-11,2675-06,2675-07} are assigned a structure-specific temperature, e.g. a liver blood vessel has a constant temperature of \SI{37}{\celsius}.
The temperature flow in the human body consists of three constituents \cite{benzinger1969heat}: (1) Heat conduction from warm to cold tissue. (2) Cooling by blood flow \cite{2675-02}. (3) Metabolic heat generation.
The model to compare against for temperature propagation using liver ablations in this work is the modified Bioheat equation as stated in Linte et al. \cite{2675-05,2675-02}:

\begin{equation}
\label{2675-eq1_Linte}
\rho c_{p}\frac{\partial T}{\partial t} = \triangledown (K\triangledown T)+\sigma |\triangledown V|^{2}-\rho_{b}w_b c_{b}\cdot\alpha (t)(T-T_{a}).
\end{equation}

Assessment of the relevant contribution of different terms of Eq.  \ref{2675-eq1_Linte} vs. \textit{in vitro} experiments \cite{2675-05} has lead us to a more robust simplified simulation equation used in this work for liver tumor ablation:

\begin{equation}
\label{2675-eq1_Kath}
\rho c_{p}\frac{\partial T}{\partial t}=\triangledown(K \triangledown T)+w_{b}c_{b}(T_{a}-T)+Q_{m}.
\end{equation}

The variables denote 
the resistive heat transfer at Linte's needle tip $\sigma |\triangledown V|^{2}$ (with $V$ as the electric potential (\si{\volt})), 
the blood density $\rho_{p}$ (\si[per-mode=fraction]{\kilogram\per\metre\tothe{3}}) and 
the tissue state coefficient $\alpha(t)$ (Sec. \ref{sec:tissState}) from Eq. \ref{2675-eq1_Linte}. Eq. \ref{2675-eq1_Kath} incorporates 
the metabolic heat generation $Q_{m}$ of liver cells (\SI[per-mode=fraction]{10714}{\watt\per\metre\tothe{3}}) \cite{2675-03}.
In both equations, $\rho$ the specific density of liver tissue (\SI[per-mode=fraction]{1079}{\kilogram\per\metre\tothe{3}}),
$c_{p}$ denotes the liver tissue heat capacity (\SI[per-mode=fraction]{3540}{\joule\per\kilogram\per\celsius}),
$t$ the time (\si{\second}),
$K$ the liver thermal conductivity (\SI[per-mode=fraction]{0,52}{\watt\per\metre\per\celsius}), 
$T$ the temperature (\si{\celsius}), 
$w_{b}$ the liver blood flow rate (\SI[per-mode=fraction]{16,687}{\kilogram\per\metre\tothe{3}\per\second}), 
$c_{b}$ the heat capacity of the blood (\SI[per-mode=fraction]{3617}{\joule\per\kilogram\per\celsius}), and
$T_{a}$ the blood temperature (\SI{37}{\celsius}). 

In contrast to Linte, in our planning system we model the needle tip or heat source with complex geometries (e.g. umbrellas), blood vessels as boundary conditions and breathing motion simulation, use a metabolic heat source $Q_m$ and neglect external cooling $Q_p$ (air) (\si[per-mode=fraction]{\watt\per\meter\tothe{3}}) from the historical Pennes formula \cite{2675-02}.
The simulation of breathing motion is optionally possible, see Sec. \ref{sec:breathing}.
Compared to Pennes \cite{2675-02}, this work ignores the
external sources spatial heat up phase $q_{p}$ in (\si[per-mode=fraction]{\watt\per\metre\tothe{3}}), because in our situation in the liver there is no constant external cooling (air) and the blood cooling is already included in
the middle term on the right side of the PDE. Moreover vs. Linte, we use a liver-specific $Q_m$ \cite{2675-03}.
Finally, Linte's tissue decay term from Eq. \ref{eq-tissState} contains a singularity regarding $c(0)$ and yields high values near $t=0$.

For 3D continuous space, the partial differential equation can be written as:

\begin{equation}
\label{2675-eq2}
\rho c_{p}\frac{\partial T}{\partial t} = K\cdot\left(\frac{\partial^2T}{\partial x^2}+\frac{\partial^2T}{\partial y^2}+\frac{\partial^2T}{\partial z^2}\right)+w_{b}c_{b}(T_{a}-T)+Q_{m}.
\end{equation}

The finite-difference method (FDM) with symmetric partial difference quotients is used to discretize the partial derivatives on a regular grid of voxels, i.e. here for $x$: 

\begin{equation}
\label{2675-eq3}
\frac{\partial^2 T_{i}}{\partial x^2}\approx\frac{T(x_{i+1})-2T(x_{i})+T(x_{i-1})}{(\vartriangle x)^{2}}.
\end{equation}

$\vartriangle x$ denotes the voxel distance between two voxels in the x-direction in millimeter. Neighbours to a voxel with index $i$ are the predecessor $i-1$ and the successor $i+1$ in the x-direction. The voxel distances $\vartriangle y$ and $\vartriangle z$ as well as indices $j$ {OliveGreen}{and $k$ are used accordingly} in the equations. Now, Eq. \ref{2675-eq2} can be transferred by substitution into 3D voxel image space $T^{n}_{i,j,k}$ for implementation:

\begin{equation}
\label{CUDAbioheat3d}
\begin{aligned}
\rho c_{p}\frac{T^{n+1}_{i,j,k}-T^{n}_{i,j,k}}{\vartriangle t} = K
\cdot  \frac{T^{n}_{i-1,j,k}-2T^{n}_{i,j,k}+T^{n}_{i+1,j,k}}{(\vartriangle x)^{2}}
+\frac{T^{n}_{i,j-1,k}-2T^{n}_{i,j,k}+T^{n}_{i,j+1,k}}{(\vartriangle y)^{2}}\\
+\frac{T^{n}_{i,j,k-1}-2T^{n}_{i,j,k}+T^{n}_{i,j,k+1}}{(\vartriangle z)^{2}})
+(w_{b_{i,j,k}}c_{b_{i,j,k}}(T^{n}_{a}-T^{n}_{i,j,k})+Q_{m_{i,j,k}})
\end{aligned}
\end{equation}

Inside a temperature image $T$, $T^{n}_{i,j,k}$ addresses the temperature value stored in a voxels at the position $i,j,k$ in image coordinates at the time iteration index $n\in N_0$. $w_{b_{i,j,k}}$, $c_{b_{i,j,k}}$ and $Q_{m_{i,j,k}}$ denote the parameters declared for Eq. {\ref{2675-eq2}} voxel-wise.

The Neumann boundary conditions for the blood vessel and needle tip voxels are ideally assumed as completely isolating, i.e. maintaining a constant temperature (\SI{37}{\celsius}). 
This is based on the reasonable assumption that any heating of the blood is immediately carried away by blood flow or constant tip heat by external energy:

\begin{equation}
\label{2675-eq6}
\frac{\partial^2 T}{\partial x^2} = 0, \text{for neighbouring voxel outside of liver (towards vessel or tip)}.
\end{equation}

\subsection{Consideration of Breathing Motion}\label{sec:breathing}

\begin{figure}[htpb!]
\centering
  \includegraphics[width=.5\textwidth]{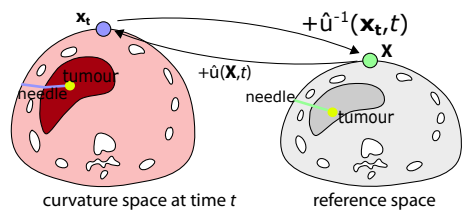}
    \caption{Time-variant diffeomorphic motion fields $\hat u$ induce a bijective relationship between reference and curvature space.}
\label{2675-fig-01}  
\end{figure}

In the 4D simulation of a liver puncture, the movement of the thorax, and here mainly the diaphragm and upper abdomen, during breathing is taken into account \cite{2675-08,2675-09}. The simulated temperature images are deformed in the reference coordinate space $\mathrm X$ with a diffeomorphic motion vector field $\hat u$ into the breathing motion's curved coordinate space $\mathrm x$ (Fig. \ref{2675-fig-01}):

\begin{equation}
\label{eq:breathing}
T\left(\mathbf{x}_{t}\right)=T_{ref}\left(\mathbf{x}_{t}+\hat{u}^{-1}(\mathbf{x}_{t},t)\right).
\end{equation}

The motion vector fields are computed from consecutive 3D CT phase images (static breathing states) taken from a breathing cycle scanned over time in a 4D CT data set. Non-linear registrations are calculated between the phases resulting in inter-phase compensation motion vector fields $u_{phase}$. In a keyframe approach, consecutive vector fields $u_{phase}$ and $u_{phase+1}$ serve to interpolate the motion of the cycle as scanned. In a linear regression approach, the vector fields (regressand) and a breathing surrogate signal (regressor, e.g. breathing volume) yield a time variant vector field $\hat{u}(\mathbf{x}_{t},t)$, optionally with surrogate induced irregular breathing motion \cite{mastmeyer2017interpatient,mastmeyer2018population}.

\subsection{Implementation}

The Bioheat equation is a computationally intensive formula. Thus, the Bioheat equation is implemented using Nvidia CUDA and runs massively parallelized on a GPU with thousands of computing cores \cite{2675-21}.

The used subsampled and segmented CT data set (dimensions 256$^2$x234, spacing \SI{1,552}{\milli\metre}$^2$ x\SI{2}{\milli\metre}) is converted into a first thermal image ($t=n=0$) for Alg.~\ref{CUDAalg}. The voxels' label values of the segmented data set are replaced by the temperatures of the denoted tissue \cite{2675-03}.

\begin{figure}[htpb!]
    \centering
    \includegraphics[width=.5\textwidth]{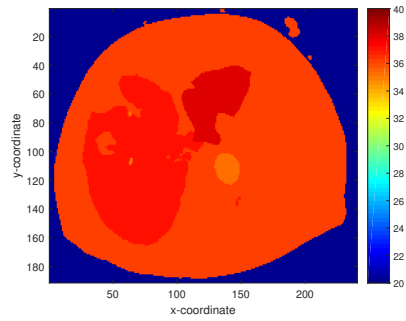}
    \caption{The initial ($t=0$) temperature image (color coded \si{\celsius} right)\cite{2675-23}.}
    \label{fig:startTemperature}
\end{figure}

Upon this start image (Fig. \ref{fig:startTemperature}), the heat equation Eq. \ref{CUDAbioheat3d} is calculated and yields a temperature update value per voxel. In our VR simulator, an image double buffer \cite{priem1996apparatus} is used to avoid rendering artifacts from unfinished calculations.
Double buffering switches back and forth between two buffers using memory pointers, see Alg. \ref{DBalg}.

\subsection{Evaluation Methods}

The temperature propagation simulation proposed here has been compared with \textit{in vitro} measurements on real tissue and the simulation by Linte et al. \cite{2675-05,2675-18}. To compare in a fair manner, the model of this work is calculated on the same high resolution \SI{0,5}{\milli\meter\tothe{3}} voxels. The \SI{2,5}{\milli\meter} geometry of the needle tip is modelled congruent to Linte et al. \cite{2675-05} as a voxel hemisphere by five voxels in xy-direction and three voxels in z-direction. Temperatures of \SI{90}{\celsius} and \SI{60}{\celsius} are applied. Simulation time steps of \SI{0,1}{\second} were selected. 

\begin{figure}[htpb!]
    \centering
    \includegraphics[width=.5\textwidth]{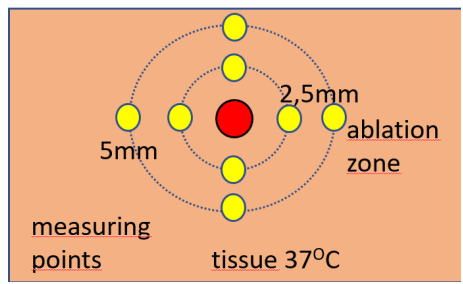}
    \caption{Scheme of the simulated temperature measurements: Around the tip (red), several temperatures are measured (yellow).}
    \label{measurement}
\end{figure}

Linte et al. \cite{2675-05,2675-18} conducted clinically relevant \textit{in vitro} experiments. Alas, they were not reproducible by us to create more data samples, mainly either because the algorithm was not reproducable or made available from the authors work.
The gold-standard for the comparison experiments in Linte's \textit{in vitro} experiments results from repeated measurements using fresh tissue samples. Symmetrically arranged temperature probes were placed in 2.5 mm and 5 mm distance from the heat source. Four according reference mean curves result for two distances vs. two ablation temperatures (\SI{60}{\celsius} and \SI{90}{\celsius}) \cite{2675-18}.
Our according \textit{in silico} simulation measurement points are shown in Fig. \ref{measurement}.

The "thread per block configuration" performance of the method is measured in frames per second (FPS). For measurements, a consumer Nvidia GTX 1080 GPU is used with 8 Gb memory and 2560 CUDA cores.

\section{Results}

\begin{figure}[htpb!]
\centering
\label{curve90C25mm}
\includegraphics[width=.5\textwidth]{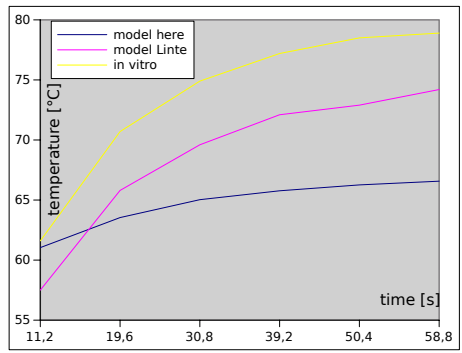}
\caption{\label{25mm90C} Temperatures over time (\SI{90}{\celsius}, \SI{2.5}{\milli\meter}): Tendential underestimate of the cell death zone in both models \cite{2675-05}. NB: Simulations are w/o standard deviation.}
\end{figure}

\begin{figure}[htpb!]
\centering
\label{curve60C25mm}
\includegraphics[width=.5\textwidth]{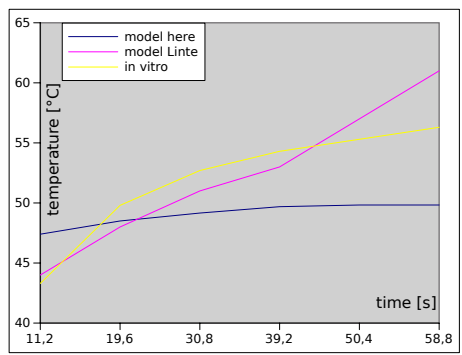}
\caption{ \label{25mm60C} Temperatures over time (\SI{60}{\celsius}, \SI{2.5}{\milli\meter}): harmful overestimation of the temperature in Linte et al. \cite{2675-05}. NB: Simulations are w/o standard deviation.}
\end{figure}

\begin{figure}[htpb!]
\centering
\label{curve60C5mm}
\includegraphics[width=.5\textwidth]{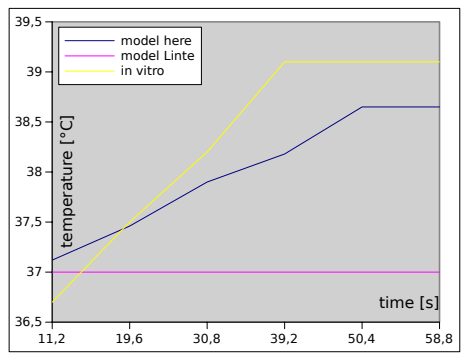}
\caption{\label{5mm60C}Temperatures over time (\SI{60}{\celsius}, \SI{5}{\milli\meter}): harmful no heating prediction by Linte et al. \cite{2675-18}, conservative underestimation in model of this work. NB: Simulations are w/o standard deviation.}
\end{figure}

\begin{figure*}[htpb!]
\begin{centering}
\begin{subfigure}[t]{0.3\textwidth}
\includegraphics[width=\textwidth]{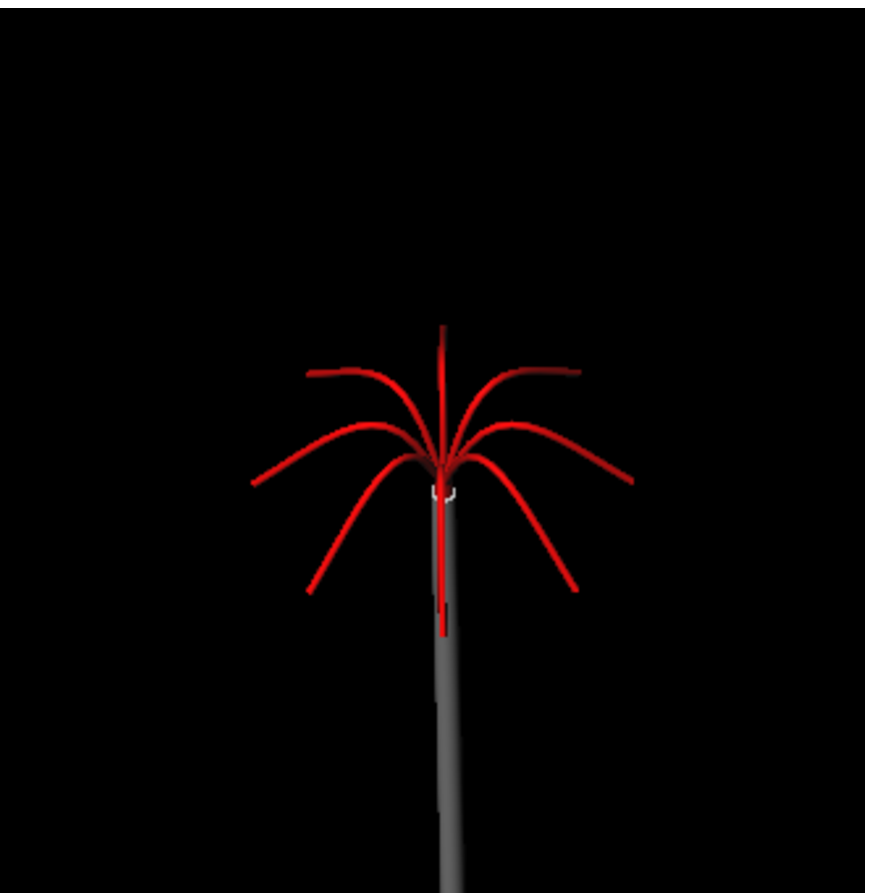}
\caption{Needle model with 8 wires}
\end{subfigure}
\begin{subfigure}[t]{0.3\textwidth}
\includegraphics[width=\textwidth]{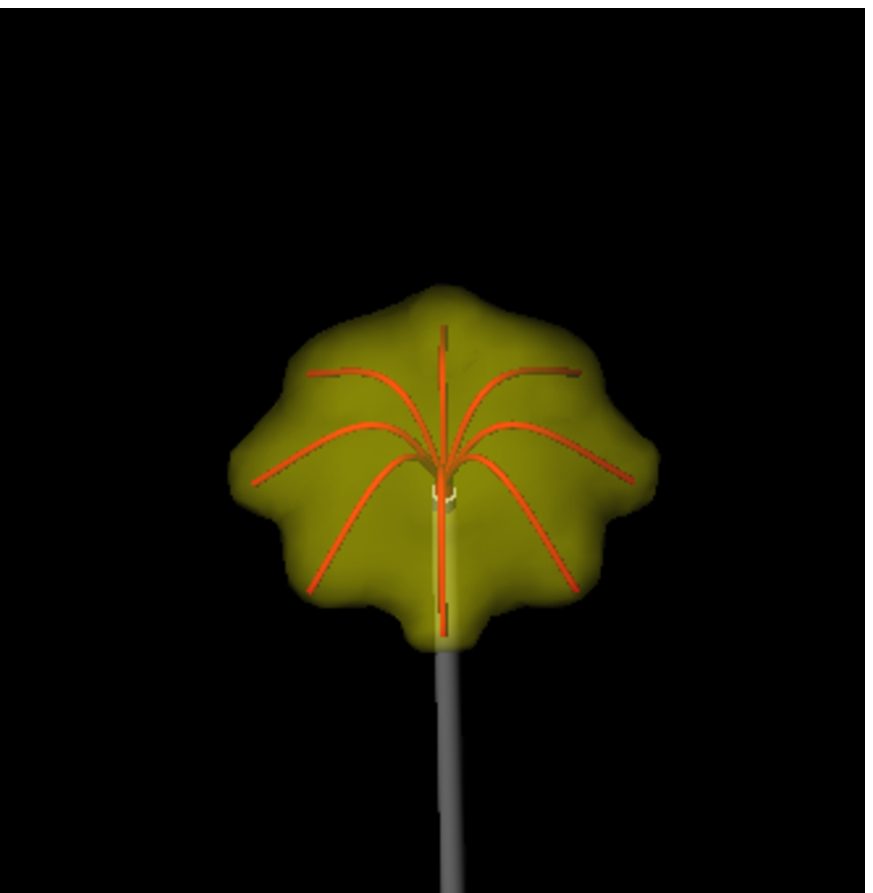}
\caption{ Propagation after \SI{1}{\second}}
\end{subfigure}
\begin{subfigure}[t]{0.3\textwidth}
\includegraphics[width=\textwidth]{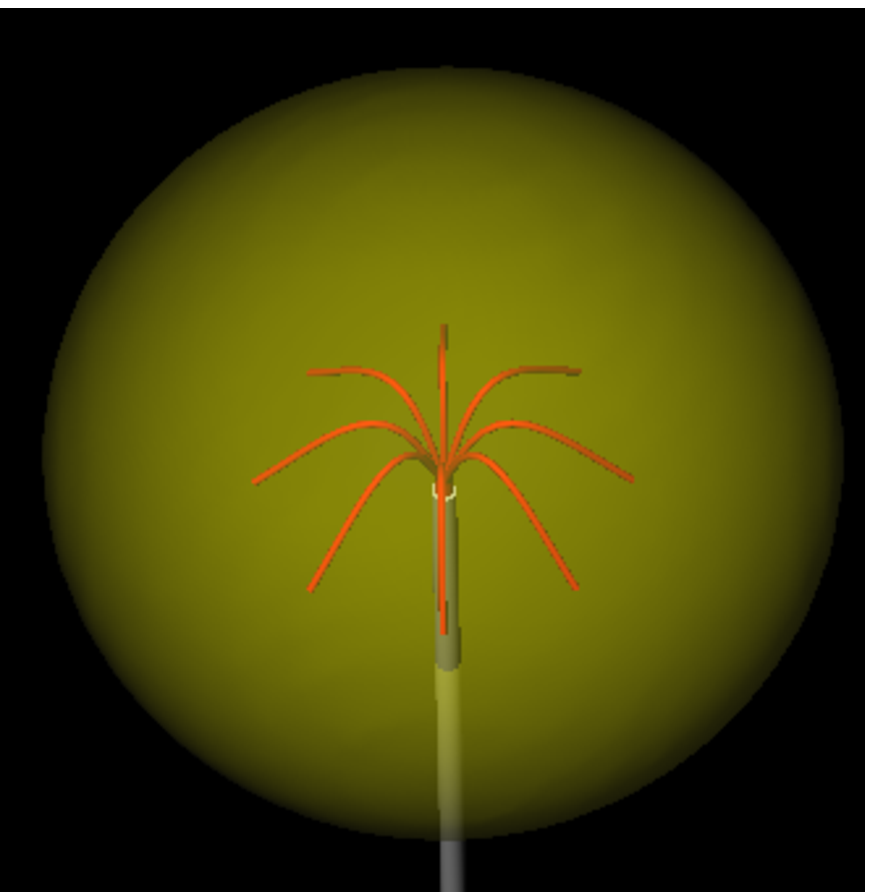}
\caption{ Propagation after \SI{58}{\second}}
\end{subfigure}
\caption{Visualization of the \SI{42.5}{\celsius} convergent spherical tissue death zone (transparent yellow) with complex needle geometry.} 
\label{needlescreen}
\end{centering}
\end{figure*}

First the quantitative results (Figs. \ref{25mm90C}, \ref{5mm90C}, \ref{25mm60C}, \ref{5mm60C}), then qualitative simulation examples near a blood vessel and using a complex ablation electrode geometry are shown (Figs. \ref{2675-fig-03}, \ref{needlescreen}).

\subsection{Simulation of Heat Propagation}

\begin{figure}[htpb!]
\centering
\label{curve5mm90C}
\includegraphics[width=.5\textwidth]{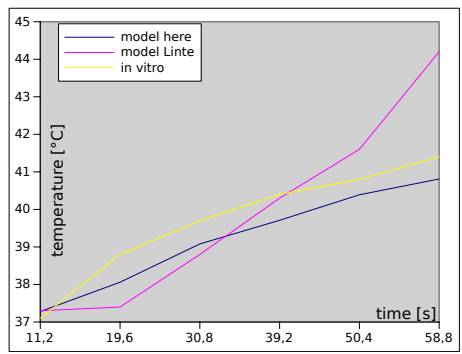}
\caption{ \label{5mm90C} Temperatures over time (\SI{90}{\celsius}, \SI{5}{\milli\meter}): harmful overestimation of the cell death zone in Linte et al. \cite{2675-05}. NB: Simulations are w/o standard deviation.}
\end{figure}

\begin{figure*}[htpb!]
\begin{subfigure}[t]{0.5\textwidth}
\includegraphics[width=\textwidth]{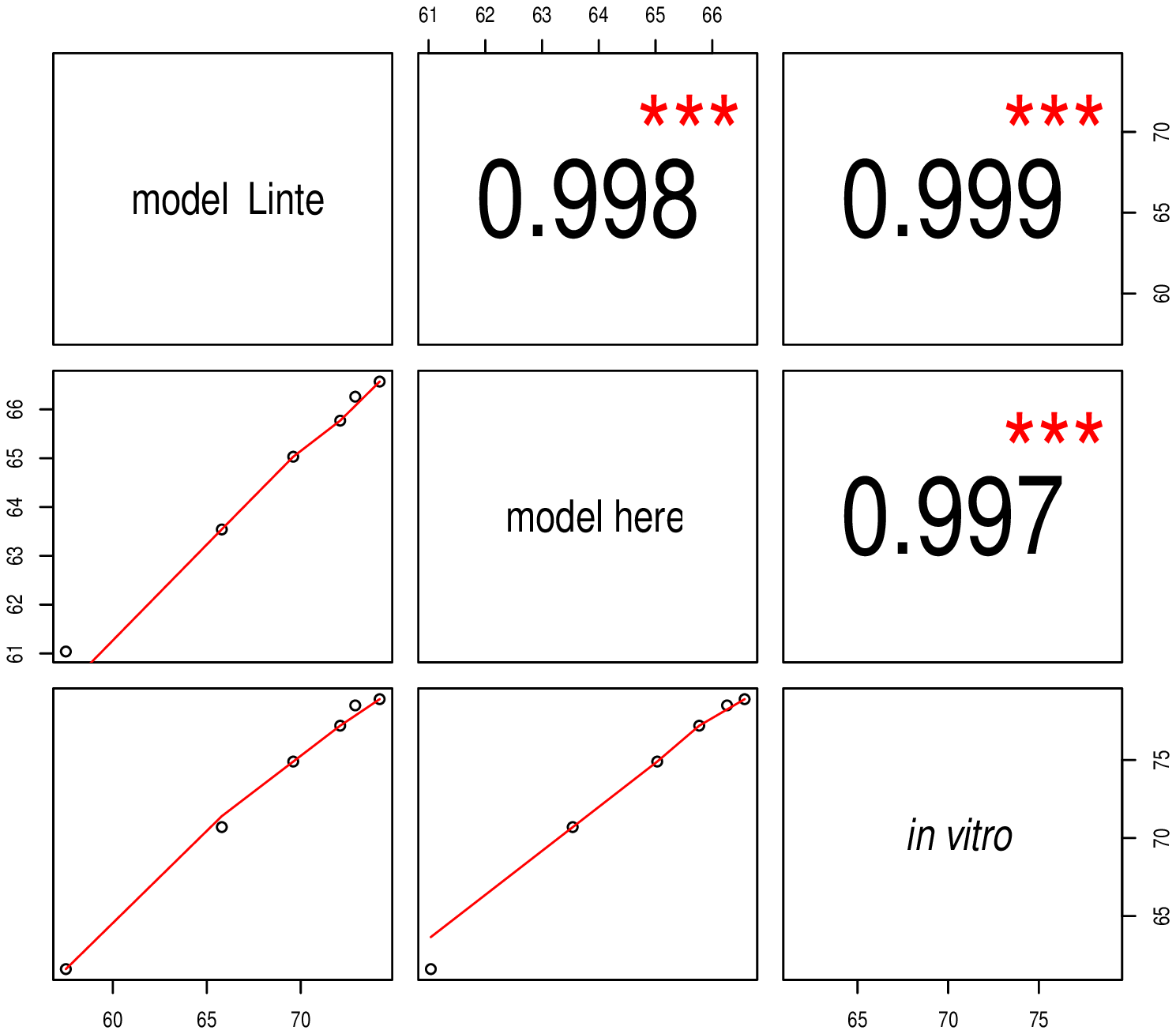}
\caption{\SI{90}{\celsius}, \SI{2.5}{\milli\meter}}
\end{subfigure}
\begin{subfigure}[t]{0.5\textwidth}
\includegraphics[width=\textwidth]{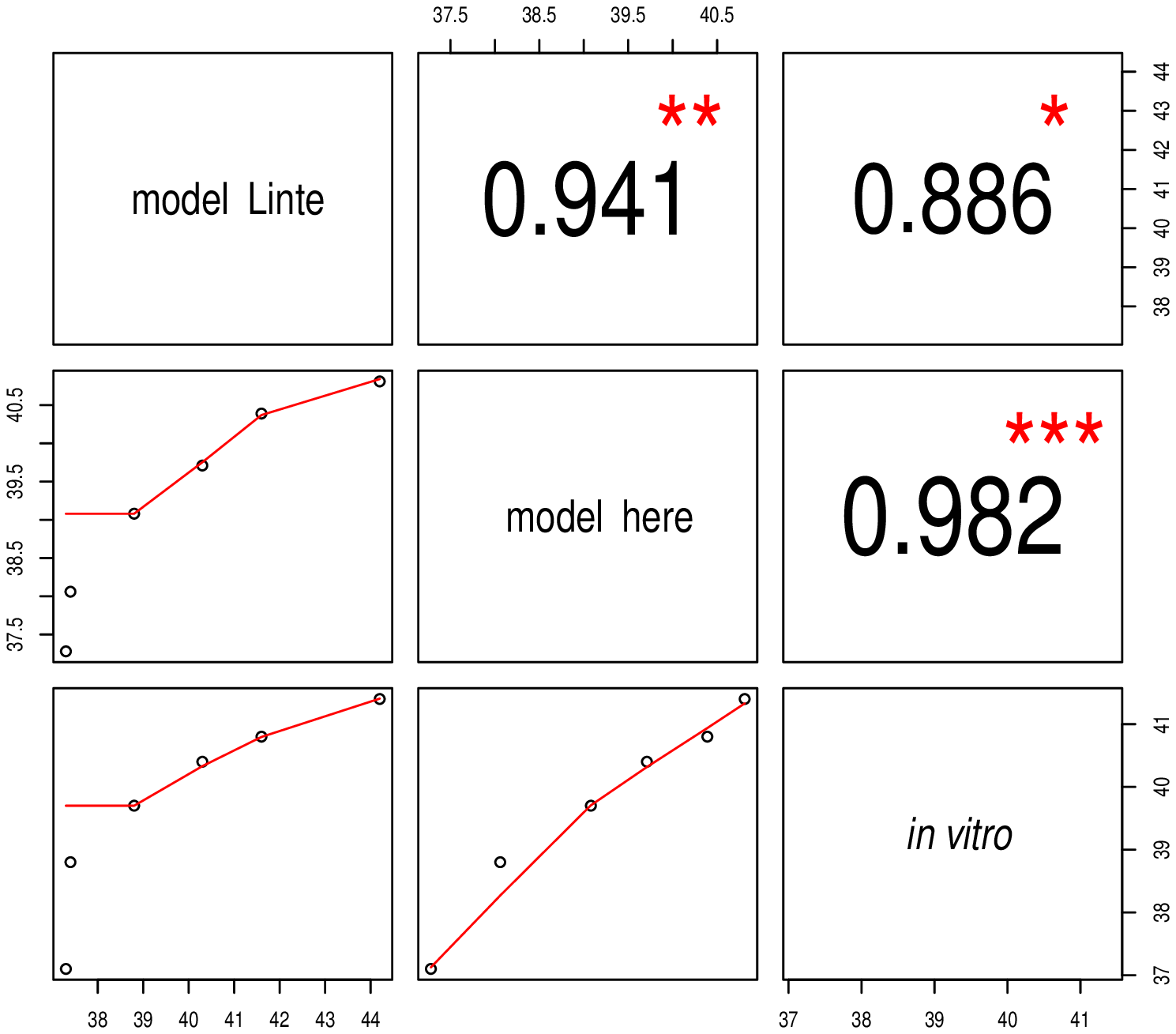}
\caption{\SI{90}{\celsius}, \SI{5}{\milli\meter}}
\end{subfigure}
\begin{subfigure}[b]{0.5\textwidth}
\includegraphics[width=\textwidth]{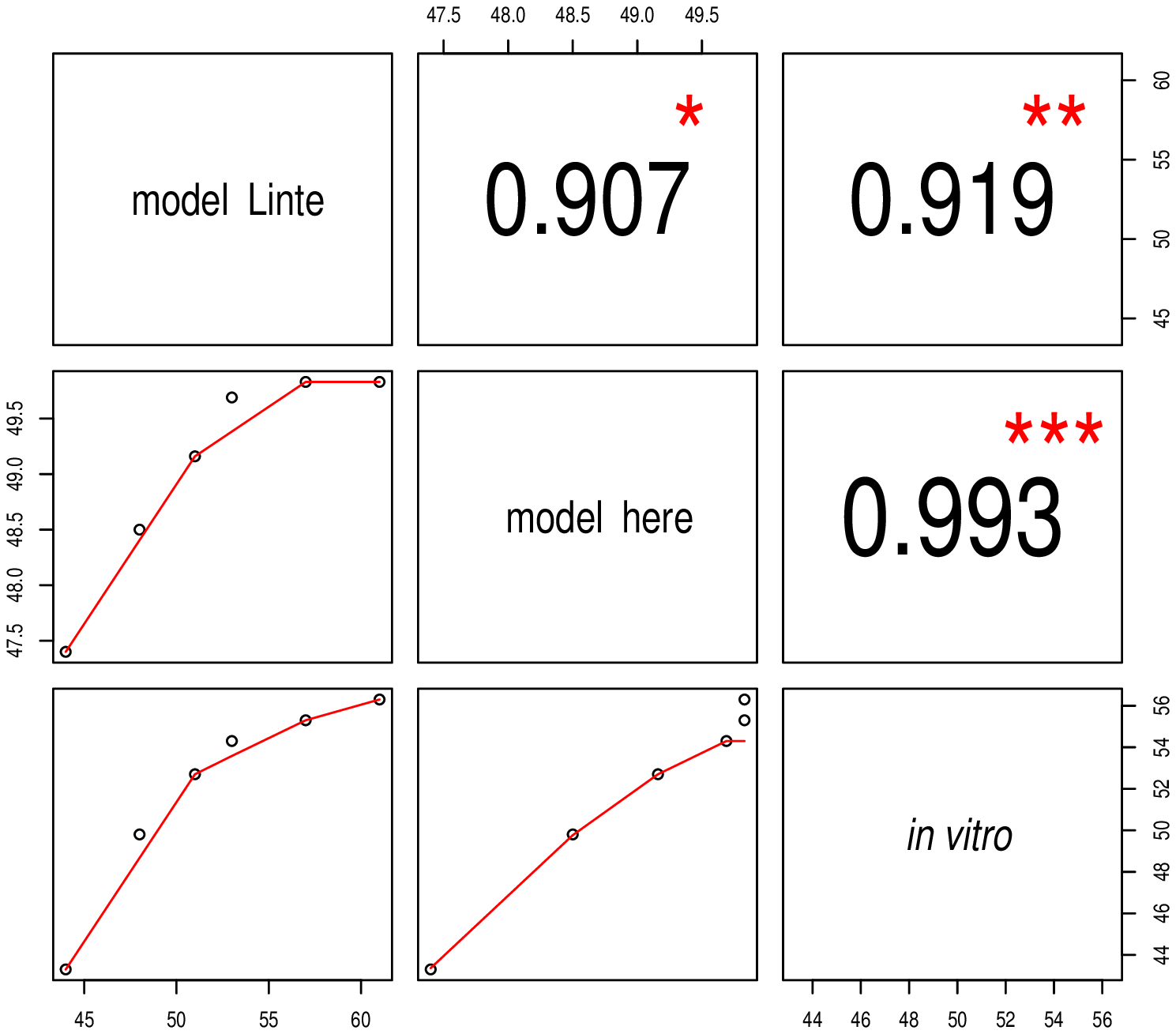}
\caption{\SI{60}{\celsius}, \SI{2.5}{\milli\meter}}
\end{subfigure}
\begin{subfigure}[b]{0.5\textwidth}
\includegraphics[width=\textwidth]{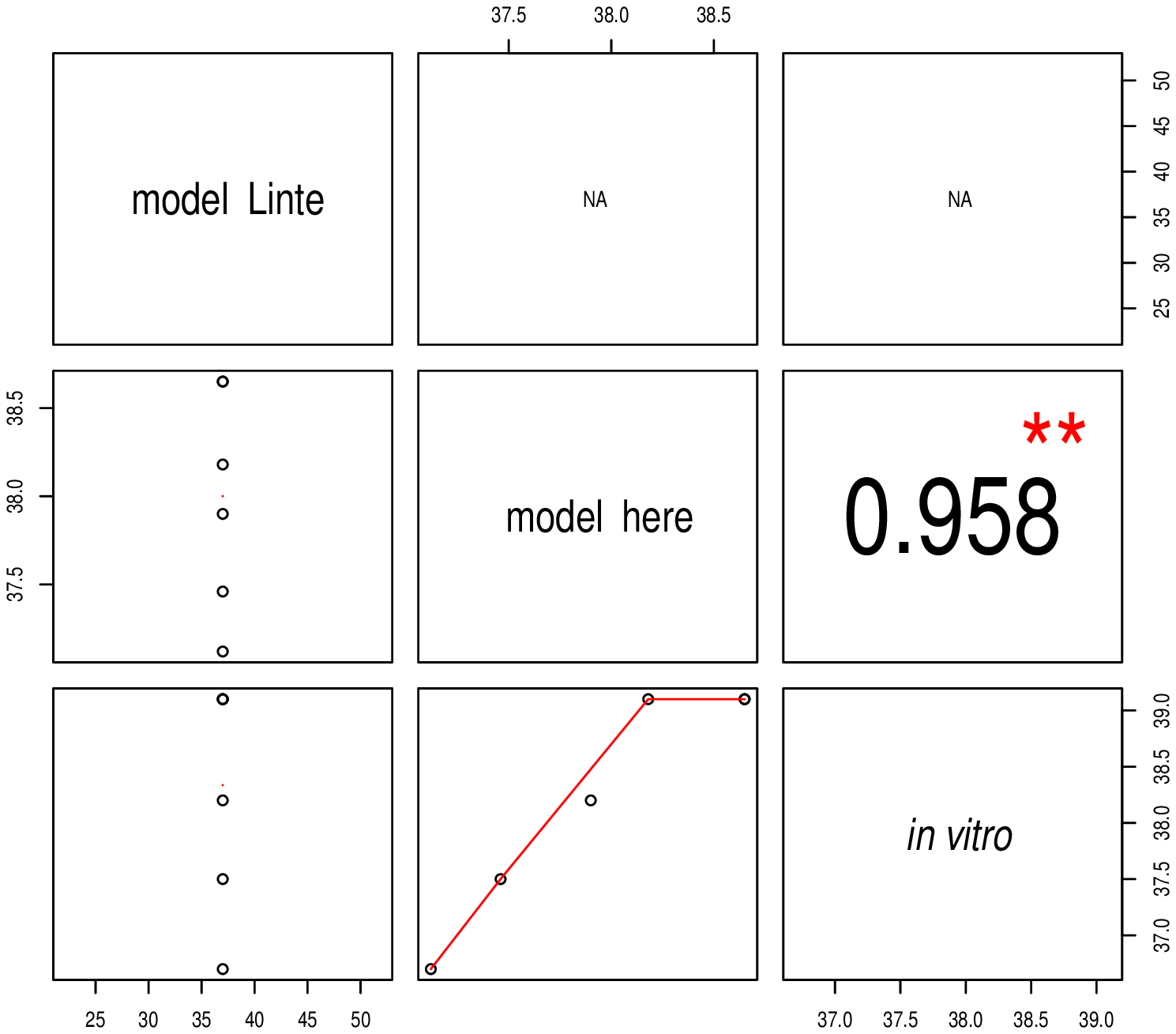}
\caption{\SI{60}{\celsius}, \SI{5}{\milli\meter}}
\end{subfigure}
\caption{Pearson correlations (*: $p<0.05$; **: $p<0.01$; ***: $p<0.001$): In 3 of 4 experiments, the model of this work shows a statistically relevant better correlation to the \textit{in vitro} temperatures (b, c, d).}
\label{pearson}
\end{figure*}

The experiment with a radius of \SI{5}{\milli\metre} and an ablation heat of \SI{90}{\celsius} is a key experiment, where Linte's method shows significant under-planning in terms of ablation time (vice versa: ablation zone predicted too big). His method is unsafe and would harm the patient. It is important to note, simulation results are absolutely reproducible without standard deviations. The temperatures measured \textit{in vitro} (gold standard) and our simulation did not exceed the tissue death threshold of \SI{42,5}{\celsius} (Fig. \ref{5mm90C}) after one minute.
With smaller errors the model presented here as well as the \textit{in vitro} determined temperatures are below the threshold of \SI{42,5}{\celsius}. The model of Linte \cite{2675-05} is on average farther away from the gold standard with a larger range of fluctuation of SD= \SI{0.99}{\celsius}. 
The model from this paper delineates the temperature distribution for dying cells more accurately, robustly and safer than the comparative model \cite{2675-05} as no failures (Fig. \ref{5mm60C}) or overestimations are observed (Figs. \ref{5mm90C}, \ref{25mm60C}). 
Statistically, this work's model shows a systematically more similar, stable and theretically sound temperature curve to the gold standard.
Summing up, in the key experiments with \SI{90}{\celsius} or \SI{60}{\celsius} and \SI{5}{\milli\meter} radius (Fig. \ref{5mm90C}), Linte critically or totally fails. 
Overall, our model results are similar to the \textit{in vitro} temperatures, which reflects in significantly high ($p<0.005$) Pearson correlation coefficients (Fig. \ref{pearson}).

\begin{figure}[htpb!]
\centering
\begin{subfigure}[b]{0.475\textwidth}
    \centering
    \includegraphics[height=3cm]{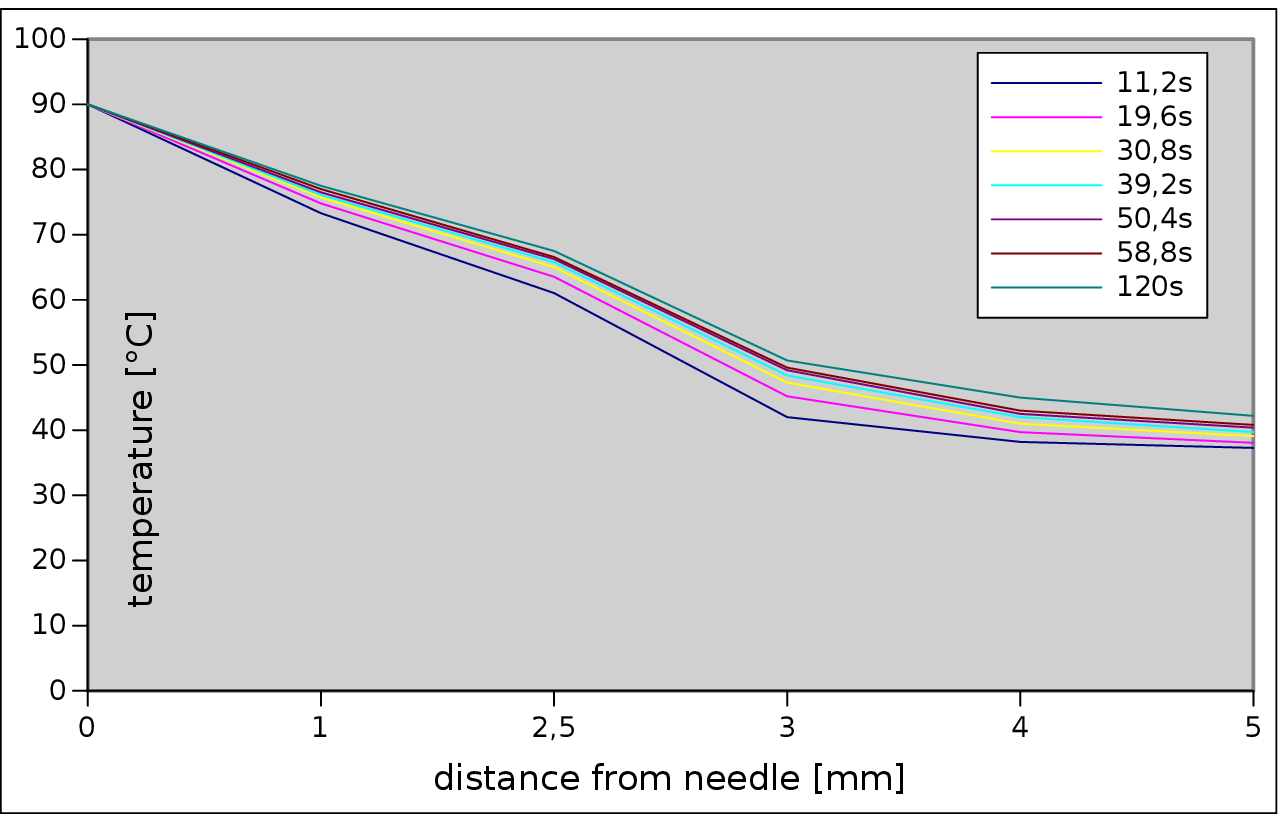}
    \caption{Distance characteristic}
    \label{distance}
\end{subfigure}
\begin{subfigure}[b]{0.475\textwidth}
    \centering
    \includegraphics[height=3cm]{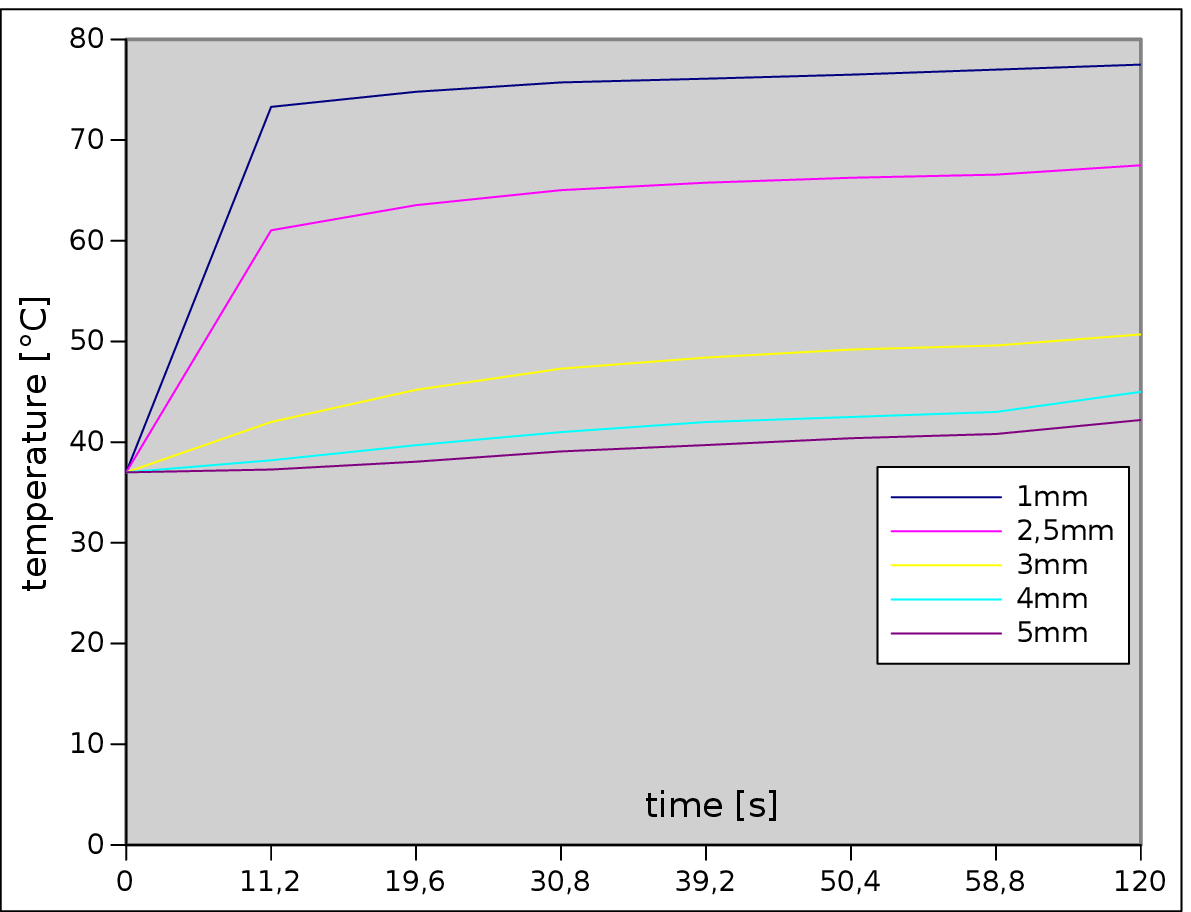}
    \caption{Time characteristic}
    \label{time}
\end{subfigure}
\caption{Simulation results over 120 sec. that compare consistently with the theoretical results in Figs. \ref{anaSpace} and \ref{anaTime}.}
\end{figure}

The simulation of the model of this work proves a theoretically sound monotonically converging heating process (Figs. \ref{distance}, \ref{time}).

\begin{figure}[htpb!]
\begin{centering}
\begin{subfigure}[t]{0.3\textwidth}
\includegraphics[width=\textwidth,height =3.135cm]{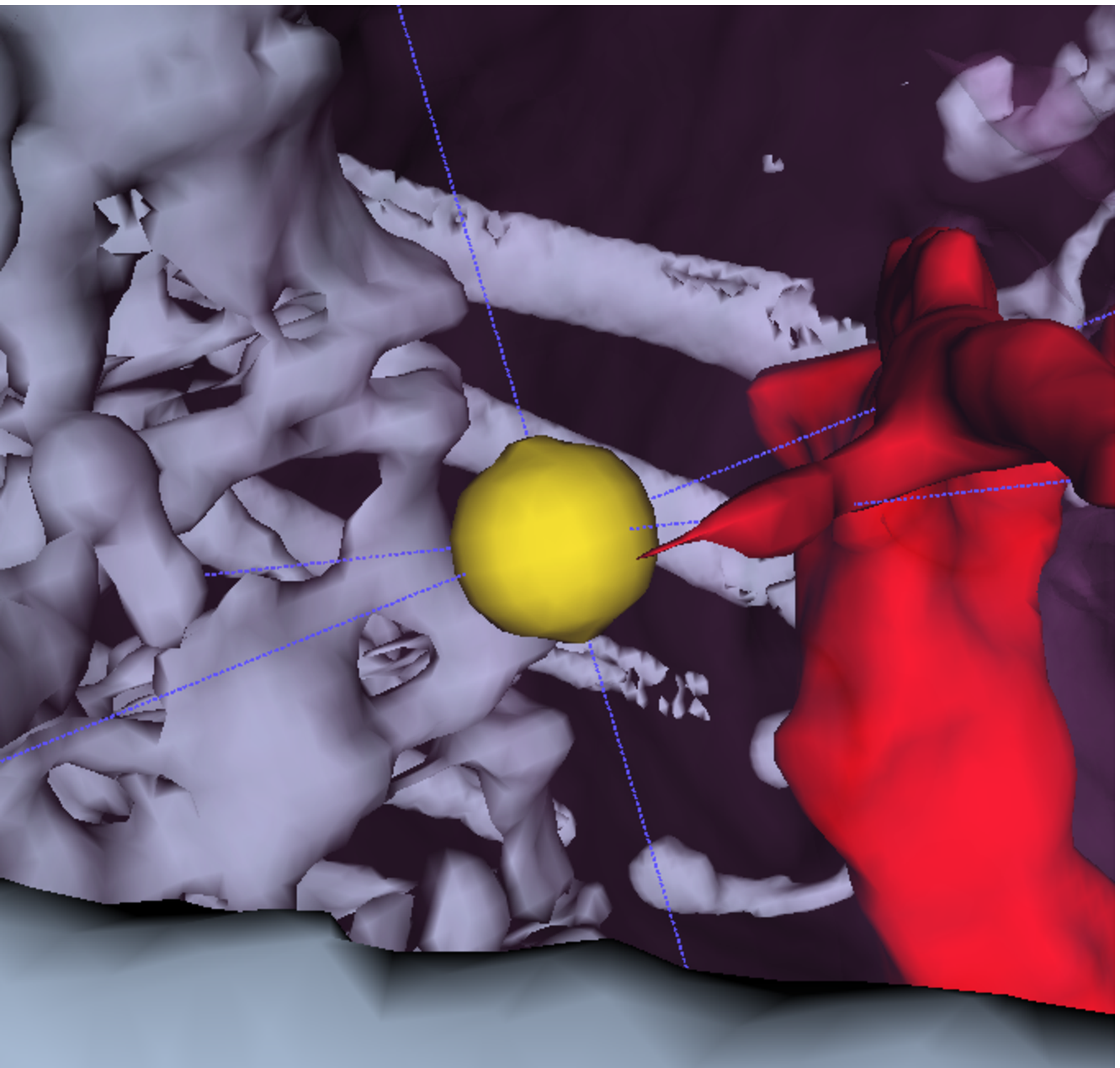}
\caption{Ablation distal}
\end{subfigure}
\begin{subfigure}[t]{0.3\textwidth}
\includegraphics[width=\textwidth,height =3.135cm]{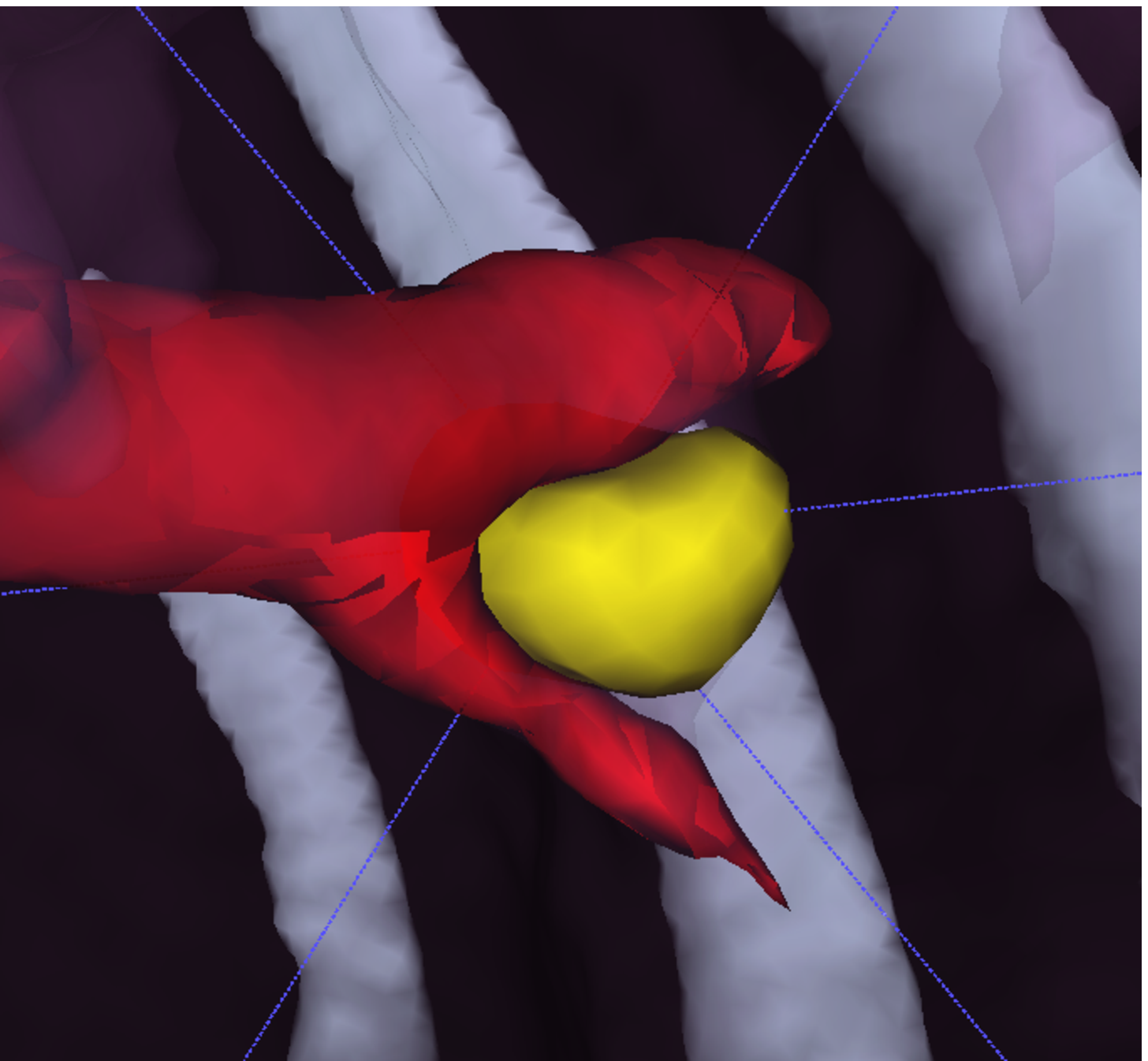}
\caption{Ablation proximal}
\end{subfigure}
\begin{subfigure}[t]{0.3\textwidth}
\includegraphics[width=\textwidth,height =3.135cm]{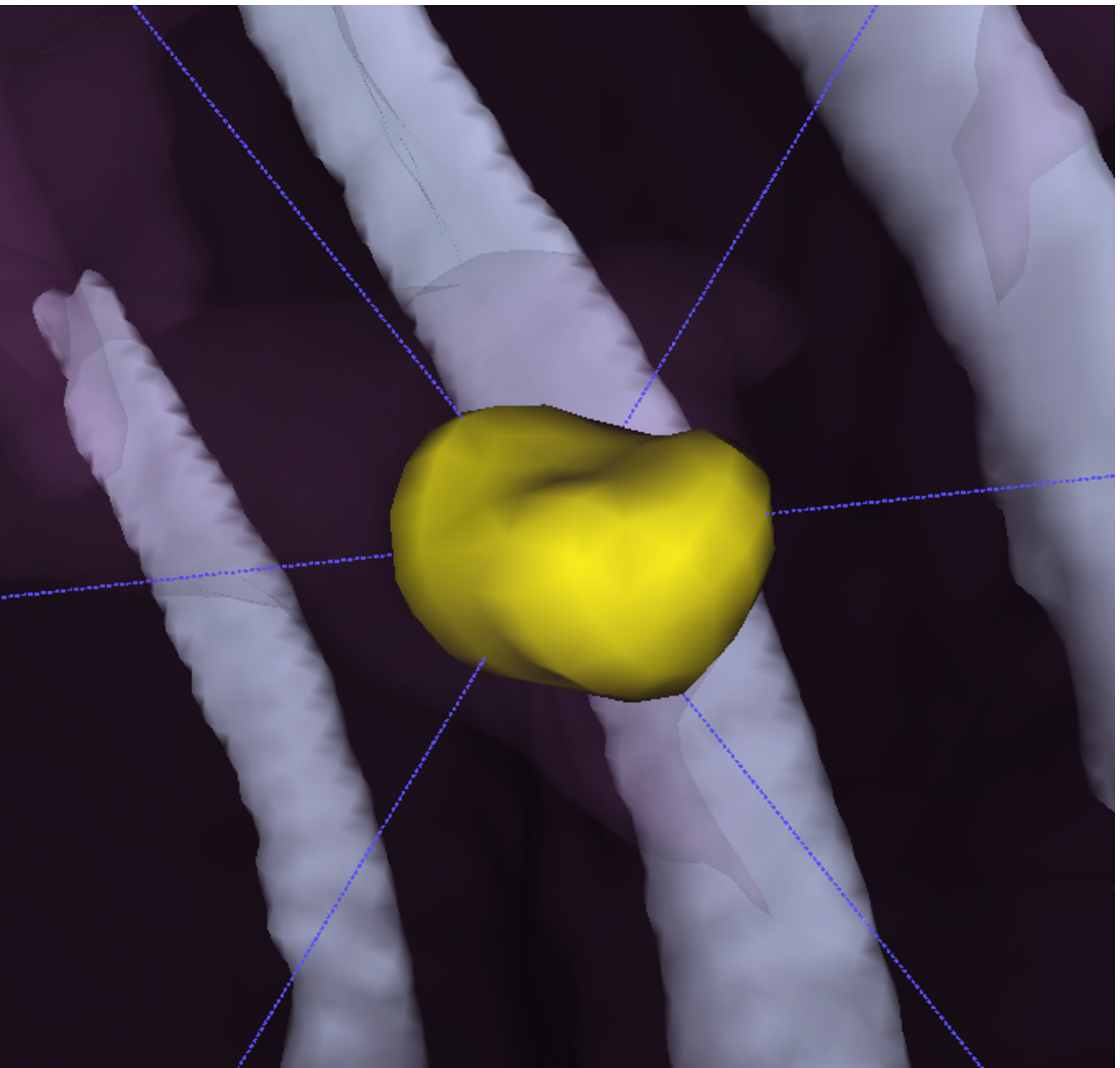}
\caption{Abl. prox. without vessel}
\end{subfigure}
\caption[Comparison of the temperature zones >\SI{42,5}{\celsius} (yellow) with different distance to hepatic artery]{(a) Spherical death zone (\SI{42,5}{\celsius}, yellow) without cooling artery. (b, c) Cooling creates an aspherical zone.}
\label{2675-fig-03}
\end{centering}
\end{figure}

In the qualitative results and without the influence of a tissue border, the propagation of the cell death zone converges to a spherical shape ((Fig. \ref{2675-fig-03}a)). If another colder tissue structure influences the heat zone, its shape adapts to the boundary of the other tissue due to enhanced cooling (Fig. \ref{2675-fig-03}b, c).
Finally, a consideration of the simulation of complex needle geometries such as umbrella needle shapes (Fig. \ref{needlescreen}) is interesting for the treatment of larger tumors ($r>$\SI{5}{\milli\meter}). 
First the tissue between the individual hot wires heats up \textit{in vitro} and then the heat zone converges to a spherical shape.
The first supplied movie demonstrates the development of the ablation zone over one minute\footnote{\url{https://goo.gl/6VPxSU}}. Second in the context of the 4D-VR simulator with simulated breathing motion according to Eq. \ref{eq:breathing}, a tumor lesion with a yellow fringe is shown in the top left simulated fluoroscopy view \footnote{\url{https://bit.ly/2HamKtO}}. In the movie footage, the yellow border of the ablation zone corresponds to \SI{42.5}{\celsius} and red denotes \SI{90}{\celsius}.

\subsection{Rendering-Performance}

All kernel configurations (Tab. \ref{2675-runtimes}) achieve over 24 frames per second (FPS) by far, making it possible to achieve a smooth display needed in moving images display. All configurations could be used for {real-time} simulation of temperature propagation. The approximately $12 \%$ faster runtime of the configuration with $64 \cdot 3 \cdot 3$ threads per block is due to the better use of the shared memory of CUDA in the calculation of the finite differences.

\section{Discussion}

A highly efficient, accurate and safe planning method for ablation prediction is proposed in this work.
It has been shown that the model of this work achieves qualitatively plausible and quantitatively robust and safe results with different ablation temperatures and in terms of heat distribution. We take into account the influence of tissue boundaries (e.g. blood vessel cooling) and complex needle tip geometries. Unfortunately, we do neither have access to lab equipment or more \textit{in vitro/in silico} data for the ablations nor an implementation of Linte's algorithm.
Summing up the quantitative evaluation, the model from this work predicts the temperature distribution for dying cells more accurately, robustly and reliably than the comparative model \cite{2675-05}. With smaller curvature errors, the model presented here as well as the \textit{in vitro} determined temperatures are below the threshold of \SI{42.5}{\celsius} (Fig. \ref{5mm90C}) in a key experiment and show a systematically more similar temperature curve vs. the gold standard and most importantly vs. theoretical curves. In the studies, the model of this work shows a conservative underestimation of temperature, but in contrast to the comparison work it shows no failures, overestimates and irregular fluctuations around the gold standard while being very near to our elaborated theoretical {characteristics.}

The presented model simulated single frames with a far over-fulfilling frame rate (FPS$>$24 Hz, x23) for our VR planning simulator \cite{2675-13} and in general.
In contrast to Linte et al. \cite{2675-05}, the tissue state coefficient and thermal resistance are not considered, but the metabolic heat generation $Q_m$ of the liver is. The heat zones from this work are consistently closer to reality, i.e. the radius of the heat zone is more accurate and safe preventing tumor recurrence. The systematic underestimation in Fig. \ref{25mm90C} means a more conservative planning and is safer than an overestimation, which is critical to the patient (tumor recurrence). In only this experiment, the comparison method \cite{2675-05} won, but lost in the other three challenges for temperature (ablation zone) overestimation (Figs. \ref{5mm90C}, \ref{25mm60C}) or flat-line (Fig. \ref{5mm60C}). Note again, simulation result differences are absolutely reproducible with no standard deviations, i.e. they are highly significant per se - even if the mean gold standard curve is with statistical deviations.

The theoretical monotonically convergence characteristic is hurt in Linte's method. In RFA planning, overestimation of the heat zone can be deadly by leaving residual tumor cells. We attribute this advantage of our method to the dubious $\alpha$-term used by Linte depriving blood cooling too early. Additionally, several questions are not answered in his work: How is the cell concentration determined at the beginning of the ablation ($t=0$), what exactly is the formula for the function $c(t)$, and how is the singularity avoided. These open issues are the strongest argument to use our method, which is transparently presented here. We were not able to get the source code from Linte or re-implement his method due to the imprecise method descriptions. We were left to compare against his provided experimental and simulation results. In the light of absolutely reproducible mathematical simulations, our method is a significantly better option. The mean comparison gold standard from Linte's experiments results from multiple experiments.
Our future research is the clarification what factors are represented in the experimental (linear) error correction terms to make our model even better and later come up with a fully theoretical approach once the remaining error factors are identified - e.g. we could use a better tissue decay term with a clear estimator $c(t)$ for healthy cell concentration.
In future research, new \textit{in-vitro} experiments with significant sample size are planned by us without necessarily comparing to Linte et al.

\begin{acknowledgements}
Funding: DFG: MA 6791/1-1; Nvidia GPU Grant 2018 (Mastmeyer).
\end{acknowledgements}

\bibliographystyle{spmpsci}      

\begin{thebibliography}{10}
\providecommand{\url}[1]{{#1}}
\providecommand{\urlprefix}{URL }
\expandafter\ifx\csname urlstyle\endcsname\relax
  \providecommand{\doi}[1]{DOI~\discretionary{}{}{}#1}\else
  \providecommand{\doi}{DOI~\discretionary{}{}{}\begingroup
  \urlstyle{rm}\Url}\fi

\bibitem{benzinger1969heat}
Benzinger, T.H.: Heat regulation: homeostasis of central temperature in man.
\newblock Physiological reviews \textbf{49}(4), 671--759 (1969)

\bibitem{chen2015efficacy}
Chen, Y.H., Lin, H., Xie, C.L., Zhang, X.T., Li, Y.G.: Efficacy comparison
  between cryoablation and radiofrequency ablation for patients with
  cavotricuspid valve isthmus dependent atrial flutter: a meta-analysis.
\newblock Scientific reports \textbf{5}, 10910 (2015)

\bibitem{2675-01}
Curley, S.: Radiofrequency ablation of malignant liver tumors.
\newblock Annals of Surgical Oncology \textbf{10}(4), 338--347 (2003)

\bibitem{evans2012numerical}
Evans, G., Blackledge, J., Yardley, P.: Numerical methods for partial
  differential equations.
\newblock Springer Science \& Business Media (2012)

\bibitem{2675-16}
Fortmeier, D., Mastmeyer, A., Handels, H.: Gpu-based visualization of
  deformable volumetric soft-tissue for real-time simulation of haptic needle
  insertion.
\newblock In: Bildverarbeitung f{\"u}r die Medizin 2012, pp. 117--122.
  Springer, Berlin, Heidelberg (2012)

\bibitem{2675-17}
Fortmeier, D., Mastmeyer, A., Handels, H.: An image-based multiproxy palpation
  algorithm for patient-specific vr-simulation.
\newblock Studies in health technology and informatics pp. 107--113 (2014)

\bibitem{2675-13}
Fortmeier, D., Mastmeyer, A., Schr\"oder, J., Handels, H.: A virtual reality
  system for {PTCD} simulation using direct visuo-haptic rendering of partially
  segmented image data.
\newblock IEEE Journal of Biomedical and Health Informatics \textbf{20}(1),
  355--366 (2016)

\bibitem{2675-08}
Fortmeier, D., Wilms, M., Mastmeyer, A., Handels, H.: Direct visuo-haptic {4D}
  volume rendering using respiratory motion models.
\newblock IEEE Transactions on Haptics \textbf{8}(4), 371--383 (2015)

\bibitem{goldberg2002comparison}
Goldberg, S.N.: Comparison of techniques for image-guided ablation of focal
  liver tumors.
\newblock Radiology \textbf{223}(2), 304--307 (2002)

\bibitem{2675-03}
Hasgall, P., Di~Gennaro, F., Baumgartner, C., Neufeld, E., Lloyd, B., Gosselin,
  M., Payne, D., Klingenb{\"o}ck, A., Kuster, N.: {IT}'{IS} {D}atabase for
  thermal and electromagnetic parameters of biological tissues (2018)

\bibitem{izzo2003other}
Izzo, F.: Other thermal ablation techniques: microwave and interstitial laser
  ablation of liver tumors.
\newblock Annals of surgical oncology \textbf{10}(5), 491--497 (2003)

\bibitem{kuck2016cryoballoon}
Kuck, K.H., Brugada, J., F{\"u}rnkranz, A., Metzner, A., Ouyang, F., Chun,
  K.J., Elvan, A., Arentz, T., Bestehorn, K., Pocock, S.J., Albenque, J.P.,
  Tondo, C.: Cryoballoon or radiofrequency ablation for paroxysmal atrial
  fibrillation.
\newblock New England Journal of Medicine \textbf{374}(23), 2235--2245 (2016)

\bibitem{2675-05}
Linte, C., Camp, J., Holmes, D., Rettmann, M., Packer, D., RA, R.: Toward
  modeling of radio-frequency ablation lesions for image-guided left atrial
  fibrillation therapy: model formulation and preliminary evaluation.
\newblock Studies in health technology and informatics \textbf{184}(11),
  261--267 (2013)

\bibitem{2675-18}
Linte, C.A., Camp, J.J., Holmes, D.R., Rettmann, M.E., Robb Richard~A.",
  e.C.A., Chen, E.C.S., Berger, M.O., Moore, J.T., Holmes, D.R.: Modeling of
  radiofrequency ablation lesions for image-guided arrhythmia therapy: A
  preliminary ex vivo demonstration.
\newblock In: Augmented Environments for Computer-Assisted Interventions, pp.
  22--33. Springer Berlin Heidelberg, Berlin, Heidelberg (2013)

\bibitem{llovet2003systematic}
Llovet, J.M., Bruix, J.: Systematic review of randomized trials for
  unresectable hepatocellular carcinoma: chemoembolization improves survival.
\newblock Hepatology \textbf{37}(2), 429--442 (2003)

\bibitem{2675-14}
Mastmeyer, A., Fortmeier, D., Handels, H.: Direct haptic volume rendering in
  lumbar puncture simulation.
\newblock In: Studies in Health Technology and Informatics: Medicine Meets
  Virtual Reality 19 - MMVR 2012, \emph{Studies in Health Technology and
  Informatics}, vol. 173, pp. 280--286. IOS Press (2012)

\bibitem{2675-06}
Mastmeyer, A., Fortmeier, D., Handels, H.: Efficient patient modeling for
  visuo-haptic {VR} simulation using a generic patient atlas.
\newblock Computer Methods and Programs in Biomedicine \textbf{132}, 161--175
  (2016)

\bibitem{2675-07}
Mastmeyer, A., Fortmeier, D., Handels, H.: Random forest classification of
  large volume structures for visuo-haptic rendering in {CT} images.
\newblock In: Proc. SPIE Medical Imaging: Image Processing, vol. 9784, pp.
  97842H--1--8. International Society for Optics and Photonics (2016)

\bibitem{2675-09}
Mastmeyer, A., Fortmeier, D., Handels, H.: Evaluation of direct haptic 4d
  volume rendering of partially segmented data for liver puncture simulation.
\newblock Nature Scientific Reports \textbf{7}(1), 1--15 (2017)

\bibitem{2675-11}
Mastmeyer, A., Fortmeier, D., Maghsoudi, E., Simon, M., Handels, H.:
  Patch-based label fusion using local confidence-measures and weak
  segmentations.
\newblock In: Proc. SPIE Medical Imaging: Image Processing, pp. 86691N--1--11.
  International Society for Optics and Photonics, Orlando, USA (2013)

\bibitem{mastmeyer2015model}
Mastmeyer, A., Pernelle Guillaume, B.L., Pieper, S., Fortmeier, D., Wells, S.,
  Handels, H., Kapur, T.: Model-based catheter segmentation in mri-images.
\newblock In: International Conference on Medical Image Computing and
  Computer-Assisted Intervention -- MICCAI (2015)

\bibitem{2675-15}
Mastmeyer, A., Pernelle, G., Ma, R., Barber, L., Kapur, T.: Accurate
  model-based segmentation of gynecologic brachytherapy catheter collections in
  mri-images.
\newblock Medical image analysis \textbf{42}, 173--188 (2017)

\bibitem{2675-10}
Mastmeyer, A., Wilms, M., Fortmeier, D., Schr\"oder, J., Handels, H.: Real-time
  ultrasound simulation for training of {US}-guided needle insertion in
  breathing virtual patients.
\newblock In: Studies in Health Technology and Informatics: Medicine Meets
  Virtual Reality 22 - MMVR 2016, \emph{Studies in Health Technology and
  Informatics}, vol. 220, pp. 219--226. IOS Press (2016)

\bibitem{mastmeyer2017interpatient}
Mastmeyer, A., Wilms, M., Handels, H.: Interpatient respiratory motion model
  transfer for virtual reality simulations of liver punctures.
\newblock Journal of World Society of Computer Graphics - WSCG \textbf{25}(1),
  1--10 (2017)

\bibitem{mastmeyer2018population}
Mastmeyer, A., Wilms, M., Handels, H.: Population-based respiratory 4d motion
  atlas construction and its application for vr simulations of liver punctures.
\newblock In: SPIE Medical Imaging 2018: Image Processing, vol. 10574, p.
  1057417. International Society for Optics and Photonics (2018)

\bibitem{2675-24}
Meloni, M.F., Chiang, J., Laeseke, P.F., Dietrich, C.F., Sannino, A., Solbiati,
  M., Nocerino, E., Brace, C.L., Lee~Jr, F.T.: Microwave ablation in primary
  and secondary liver tumours: technical and clinical approaches.
\newblock International Journal of Hyperthermia \textbf{33}(1), 15--24 (2017)

\bibitem{2675-21}
Nickolls, J., Buck, I., Garland, M., Skadron, K.: Scalable parallel programming
  with cuda.
\newblock Queue \textbf{6}(2), 40--53 (2008)

\bibitem{nikfarjam2005mechanisms}
Nikfarjam, M., Muralidharan, V., Christophi, C.: Mechanisms of focal heat
  destruction of liver tumors.
\newblock Journal of Surgical Research \textbf{127}(2), 208--223 (2005)

\bibitem{2675-25}
Niu, L.Z., Li, J.L., Xu, K.C.: Percutaneous cryoablation for liver cancer.
\newblock Journal of clinical and translational hepatology \textbf{2}(3), 182
  (2014)

\bibitem{oshowo2003comparison}
Oshowo, A., Gillams, A., Harrison, E., Lees, W., Taylor, I.: Comparison of
  resection and radiofrequency ablation for treatment of solitary colorectal
  liver metastases.
\newblock British journal of surgery \textbf{90}(10), 1240--1243 (2003)

\bibitem{2675-02}
Pennes, H.: Analysis of tissue and arterial blood temperatures in the resting
  human forearm.
\newblock Journal of Applied Physiology \textbf{1}(2), 93--122 (1948)

\bibitem{priem1996apparatus}
Priem, C., Malachowsky, C., McIntyre, B., Moffat, G.: Apparatus for selecting
  frame buffers for display in a double buffered display system (1996).
\newblock US Patent 5,543,824

\bibitem{2675-20}
Shen, W., Zhang, J., Yang, F.: Modeling and numerical simulation of bioheat
  transfer and biomechanics in soft tissue.
\newblock Mathematical and Computer Modelling \textbf{41}(11-12), 1251--1265
  (2005)

\bibitem{2675-23}
Werner, J., Buse, M.: Temperature profiles with respect to inhomogeneity and
  geometry of the human body.
\newblock Journal of Applied Physiology \textbf{65}(3), 1110--1118 (1988)

\end{thebibliography}

\FloatBarrier

\newpage
\section{Figures}

\FloatBarrier

\section{Tables}
\vspace{-.75cm}
\begin{table}[htpb!]
\caption{Kernel-layout ($x\cdot y\cdot z$-Threads/block) vs. runtime for 6000 images.}
\label{2675-runtimes}
\begin{tabular*}{\textwidth}{l@{\extracolsep\fill}ccc}
\hline
Kernel conf. & Time for 6000 images (s)& Time/image (s)& FPS (Hz)\\ 
\hline
$64 \cdot 3 \cdot 3$            & 10.8                 & 0.0018       & 556\\
$256 \cdot 3 \cdot 1$           & 12.3                & 0.00205       & 488\\
$1024 \cdot 1 \cdot 1$          & 12.23                & 0.00205      & 490\\ 
\hline
\end{tabular*}
\end{table}
    
\FloatBarrier

\vspace{-.5cm}
\section{Algorithms}
\vspace{-.75cm}
\begin{algorithm}[h!]
\caption{Pseudocode Eqs. \ref{CUDAbioheat3d} and \ref{2675-eq6}.}
\label{CUDAalg}
\begin{algorithmic}
\footnotesize
\STATE inputImage = const. initial temperature image with encoded ablation tip ($n=0$)
\STATE N = number of timesteps to simulate
\STATE n = 1
\STATE bufferImage = currentInputImage
\WHILE{$n \leq N$}
    \STATE updateImage=zero(inputImage) // zero image
    \FORALL{voxel}
        \IF{voxel inside liver and voxel != (ablationTip or vessel)}
            \FORALL{dimension}
                \STATE calculate symmetrical finiteDifference@voxel[dimension] // x, y, z
            \ENDFOR
            \STATE updateImage@voxel = calc. Eq. \ref{CUDAbioheat3d} @voxel using finiteDifferences@voxel
            \STATE bufferImage@voxel += updateImage@voxel
        \ELSE
            \STATE bufferImage@voxel = inputImage@voxel // Cf. Eq. \ref{2675-eq6}
        \ENDIF
    \ENDFOR
    \STATE{ n = n + 1}
    \STATE{return bufferImage}
\ENDWHILE
\end{algorithmic}
\end{algorithm}
\begin{algorithm}[h!]
\caption{Pseudocode of the double buffer method. "\&" denotes the "reference of" operator.} 
\label{DBalg}
\begin{algorithmic}
\footnotesize
\STATE n = 1\;
\STATE buffer1 = copy(inputImage) // congruent black image\;
\STATE buffer2 = copy(inputImage) // both are the double buffer\;
\STATE outputPointer = NULL // rendering shown to the user\;
\WHILE{true}
    \IF{\&buffer1 == outputPointer}
    \STATE buffer2 = one time iteration (N=1) of Alg. \ref{CUDAalg} with buffer1
    \STATE outputPointer = \&buffer2
    \ENDIF
    \IF{\&buffer2 == outputPointer}
    \STATE buffer1 = one time iteration (N=1) of Alg. \ref{CUDAalg} with buffer2
    \STATE outputPointer = \&buffer1
    \ENDIF
\ENDWHILE
\end{algorithmic}
\end{algorithm}

\end{document}